\documentclass[lettersize,journal]{IEEEtran}
\usepackage{amsmath,amsfonts}
\usepackage{algorithmic}
\usepackage{booktabs}
\usepackage{array}
\usepackage[caption=false,font=normalsize,labelfont=sf,textfont=sf]{subfig}
\usepackage{textcomp}
\usepackage{stfloats}
\usepackage{url}
\usepackage{multirow}
\usepackage{verbatim}
\usepackage{graphicx}
\def\BibTeX{{\rm B\kern-.05em{\sc i\kern-.025em b}\kern-.08em
    T\kern-.1667em\lower.7ex\hbox{E}\kern-.125emX}}
\usepackage{balance}
\usepackage{makecell}
\usepackage{hyperref}    
\usepackage{orcidlink}   

\newcolumntype{C}[1]{>{\centering\arraybackslash}m{#1}}

\begin{document}
\bstctlcite{BSTcontrol}
\title{A Multi-Modal Intelligent U2V Channel Model for 6G Sensing-Communication Integration}
\author{
\mbox{Shuo Wang~\orcidlink{0009-0009-8943-799X},~\textit{Graduate Student Member, IEEE}}, 
\mbox{Zengrui Han~\orcidlink{0009-0009-5884-5085},~\textit{Graduate Student Member, IEEE}}, 
\mbox{Lu Bai~\orcidlink{0000-0003-1687-0863},~\textit{Member, IEEE}}, 
and \mbox{Xiang Cheng~\orcidlink{0000-0002-5943-0326},~\textit{Fellow, IEEE}}
\thanks{S.~Wang is with the School of Software, Shandong University, Jinan 250101, P. R. China, and also with the Joint SDU-NTU Centre for Artificial Intelligence Research (C-FAIR), Shandong University, Jinan, 250101, P. R. China (e-mail: shuo.wang@mail.sdu.edu.cn).}
\thanks{Z.~Han, and X.~Cheng are with the State Key Laboratory of Photonics and Communications, School of Electronics, Peking University, Beijing 100871, China (e-mail: zengruihan@stu.pku.edu.cn; xiangcheng@pku.edu.cn).}
\thanks{L.~Bai is with the Joint SDU-NTU Centre for Artificial Intelligence Research (C-FAIR), Shandong University, Jinan, 250101, P. R. China (e-mail: lubai@sdu.edu.cn).}
}

\markboth{IEEE Transactions on Wireless Communications, vol. xx, no. xx, XX 2026}
{}
\maketitle

\begin{abstract}
This paper proposes a novel UAV-to-Vehicle (U2V) channel model for sixth-generation (6G) intelligent sensing-communication integration, based on three-dimensional (3D) scatterer prediction.
To explore the mapping relationship between physical environment and electromagnetic space, a new high-fidelity mixed sensing-communication integration U2V simulation dataset under wide-lane scenarios with different vehicular traffic densities (VTDs) and UAV heights is constructed.
Based on the constructed dataset, a novel 3D Scatterer Prediction and Distribution Estimation (3D-SPADE) algorithm is proposed, which leverages LiDAR point clouds to accurately predict the spatial distribution of scatterers.
Furthermore, the clustering of scatterers and the subsequent classification into dynamic and static types are meticulously designed for highly dynamic U2V scenarios, while reducing computational complexity and improving modeling accuracy.
As LiDAR point clouds vary over time, dynamic and static clusters evolve via 3D-SPADE, enabling precise modeling of channel non-stationarity and consistency.
Simulation results demonstrate that, in the wide-lane scenario with varying VTDs and UAV heights, the proposed 3D-SPADE consistently achieves high scatterer occupancy detection performance within the voxel grid. 
In particular, under favorable configurations, recall reaches 93.26\%, and precision reaches 95.74\%, highlighting the reliability of 3D-SPADE.
Key channel statistical characteristics are simulated and analyzed. 
These characteristics from the simulation experiments are highly consistent with ray-tracing results and exhibit better agreement than with the standardized model and inconsistent model, validating the necessity of exploring the mapping relationship and the effectiveness of the proposed model.
\end{abstract}


\section{Introduction}
\IEEEPARstart{W}{ith} the global development of the low-altitude economy, the demand for uncrewed aerial vehicles (UAVs) has experienced explosive growth \cite{background_2019}. 
Meanwhile, the accelerated deployment of sixth-generation (6G) mobile communication technologies has positioned UAV-to-Vehicle (U2V) communication networks as a critical component of the Space-Air-Ground-Sea Integrated Network (SAGSIN) \cite{6G,6G_U2V, SAGSIN}. 
Additionally, wireless channels serve as the cornerstone of wireless system analysis and design \cite{goldsmith2005wireless}, making it essential to conduct comprehensive evaluations and develop advanced U2V channel models to ensure robust communication across various U2V application scenarios.  

In the field of communications, existing channel models can generally be classified into two types, i.e., stochastic models and deterministic models \cite{modelclassification}. Stochastic models generate channel parameters by statistical methods, which offer low complexity but limited accuracy due to the randomness of the parameters. Deterministic models reproduce channel characteristics based on physical radio propagation mechanisms, achieving higher accuracy at the cost of incurring significant complexity. In complex and highly dynamic U2V communication scenarios, U2V channel modeling becomes increasingly complex due to the characteristics of three-dimensional (3D) scattering environments, and the highly dynamic, arbitrary 3D motion trajectories of a UAV and vehicles, whose speeds vary unpredictably and directions differ across different vehicles. 
In this context, the inaccuracies of stochastic models are further amplified, while the complexity of deterministic models increases sharply, making it critical to explore new modeling methods that achieve a decent trade-off between accuracy and complexity.
    
Existing U2V channel modeling studies have long focused on the trade-off between accuracy and complexity.  
Many U2V channel models adopt distinct methods tailored to specific scenarios, optimized to either reduce complexity or improve accuracy based on scenario requirements. However, scenario-specific optimization often compromises their generalizability \cite{modelsurvey}. 
The authors in \cite{shaoweddoublescatter} proposed a shadowed double-scattering model to capture key non-stationary characteristics of U2V channels with reduced complexity while preserving acceptable modeling accuracy.
In contrast, \cite{maritimemodel} incorporated sea surface fluctuations and waveguide effects to capture maritime U2V propagation characteristics, achieving higher modeling accuracy at the cost of increased model complexity. 
To achieve a balance between modeling accuracy and complexity, \cite{multiUAV} proposed an array-space-time-frequency non-stationary channel model incorporating UAV 3D continuous trajectories and rotations, which is applicable to U2V scenarios with high mobility. 
The evolution of scatterers is characterized using K-Means clustering and birth–death processes to capture non-stationarity. 
However, the requirement of manually predefining the number of clusters limits its ability to model more complex and dynamic U2V environments accurately.
To address the above-mentioned problems, recent research has focused on the application of artificial intelligence (AI) in channel modeling. 
By exploiting data-driven learning and pattern recognition capabilities, AI-based methods provide a promising means to achieve a decent balance between accuracy and complexity, even in complex environments \cite{aimodel}. 
\cite{mlmodel} proposed a hybrid geometric and machine learning-based modeling approach in U2V channel modeling, enabling explicit characterization of UAV orientation and improving parameter generation. 
\cite{nonstationarymodel} employed AI algorithms to achieve joint modeling of spatial-temporal-frequency non-stationarity for scatterer clusters. 
A deep learning-based channel tracking method using Stacked Bi-directional long short-term memory (LSTM) networks is proposed in \cite{lstmmodel}, enhancing tracking performance by leveraging historical data and bidirectional structures.
\cite{pathlossmodel} proposed a neural-network-based channel modeling approach for path loss and shadowing, exploiting deep learning to capture complex propagation characteristics.
However, these studies \cite{mlmodel}-\cite{pathlossmodel} relied solely on uni-modal radio frequency (RF) communication data, treating the models as black boxes that neglect the mapping relationship between the physical environment and electromagnetic space. This omission not only limits the physical interpretability but also hampers model performance and generalization. Without this mapping, the model struggles to capture complex environmental interactions, leading to reduced accuracy and robustness in unseen scenarios.
\begin{figure*}[!t] 
    \centering 
    \includegraphics[width=0.8\textwidth]{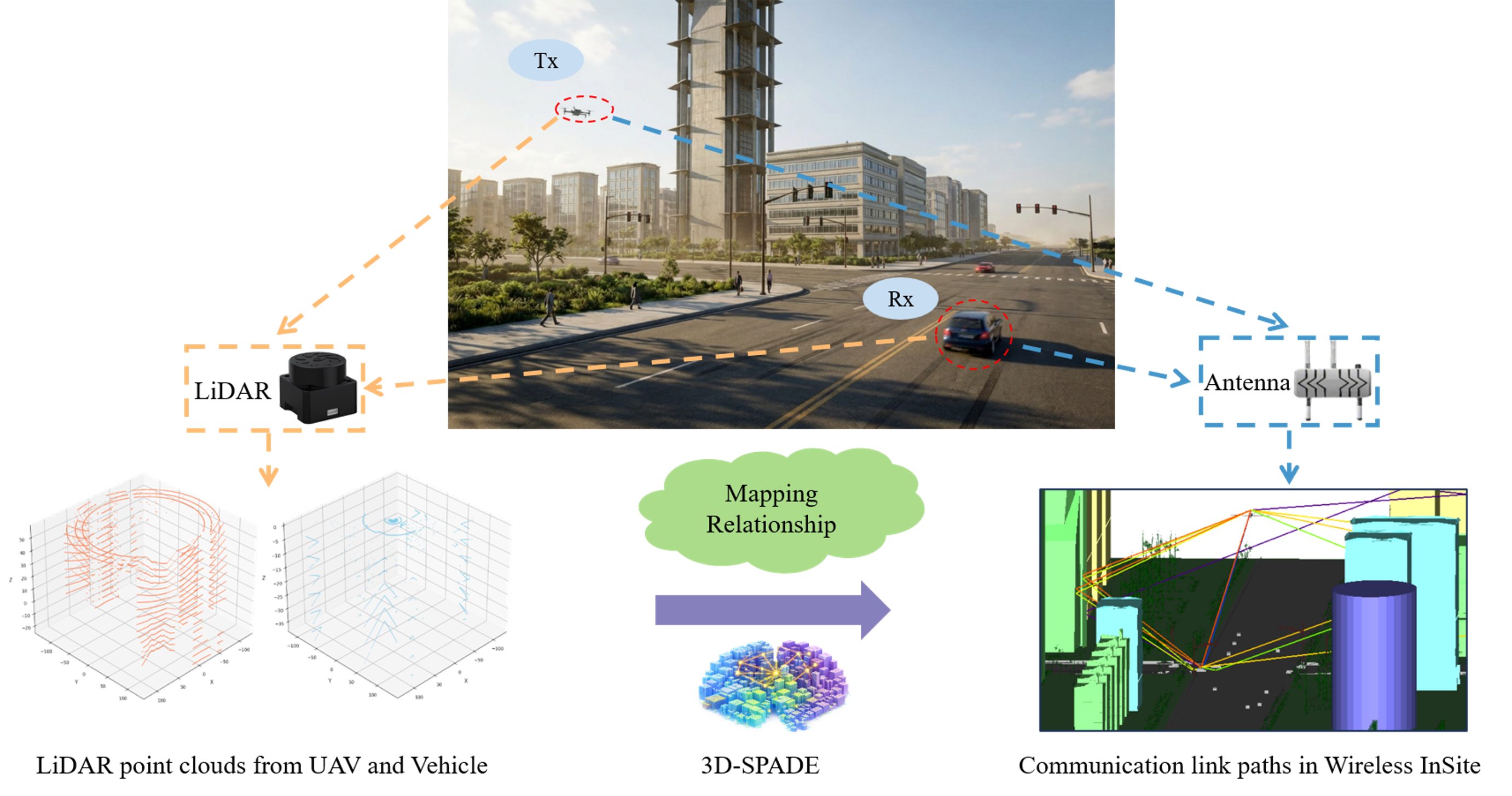} 
    \caption{The story diagram of the proposed 3D-SPADE: Exploring the mapping relationship between LiDAR point clouds and channel scatterers.} 
    \label{fig:story diagram}
\end{figure*}
To address this issue, \cite{2.6GHzmodel} and \cite{satellitemodel} employed satellite images as the core input for predicting channel path loss, providing environmental information for UAV channel modeling. However, due to limited spatial resolution and acquisition timeliness, satellite images fail to provide accurate and timely environmental information. This makes it difficult to explore the mapping relationship between physical environment and electromagnetic space, as well as to construct more accurate and lower-complexity channel models.

Fortunately, intelligent agents will be equipped with sensing devices in 6G systems, enabling richer environmental information that enhances the adaptability and intelligence of communication systems.
Inspired by human synesthesia, Cheng et al. \cite{som} proposed a novel concept, i.e., Synesthesia of Machines (SoM), which explicitly outlined the aim of intelligent multi-modal sensing-communication integration via artificial neural networks (ANNs). 
Based on the SoM, multi-modal intelligent channel modeling (MMICM) has been introduced to explore the mapping relationship between physical environment and electromagnetic space by jointly utilizing communication information and multi-modal sensory data \cite{mmicm, magazine}.
As a representative work following this paradigm, \cite{han} utilized multi-modal information from communication devices and LiDAR to explore the mapping relationship between physical environment and electromagnetic space. 
The study identified dynamic and static scatterers to construct a vehicle-to-vehicle (V2V) channel model based on the SegNet architecture, accurately capturing channel non-stationarity and consistency. 
This work demonstrates that exploiting the mapping between the physical environment and electromagnetic space can mitigate excessive reliance on data in traditional channel modeling while preserving physical interpretability, thereby achieving a reasonable trade-off between modeling accuracy and complexity.
While the method performs well in V2V scenarios, it faces three significant challenges when applied to U2V scenarios. 
First, scatterer density prediction based on the two-dimensional (2D) model is inadequate in 3D space, as it fails to capture the distinct scattering characteristics between a UAV and vehicles in 3D spatial environments. 
Second, Density-Based Spatial Clustering of Applications with Noise (DBSCAN) suffers from high computational complexity, and when applied in more complex 3D U2V environments with a large number of scatterers, the clustering speed becomes insufficient \cite{dbscan}.
Third, the 3D diverse motion trajectories of a UAV and vehicles introduce significant 3D spatio-temporal variability into the communication channel, incurring challenges for accurate channel modeling.

To address the aforementioned challenges, this paper proposes a multi-modal intelligent U2V channel model based on 3D scatterer prediction. A new high-fidelity mixed sensing-communication integration U2V simulation dataset under wide-lane scenarios with different vehicular traffic densities (VTDs) is constructed. 
Based on the constructed dataset, a 3D Scatterer Prediction and Distribution Estimation (3D-SPADE) algorithm is developed, thereby exploring the mapping relationship between physical environment and electromagnetic space, as shown in Fig.~\ref{fig:story diagram}. 
By leveraging 3D-SPADE together with clustering and cluster classification, the proposed framework enables accurate modeling of the U2V channel, effectively capturing channel non-stationarity and consistency.
The main contributions of this paper are summarized as follows.

\begin{enumerate}

\item A high-fidelity mixed sensing–communication integration U2V dataset is established for wide-lane scenarios, comprising 2 UAV flight heights and 3 vehicular traffic densities. 
For each traffic density, 10, 20, and 30 representative vehicle motion trajectories are included, enabling the characterization of diverse mobility patterns and dynamic propagation conditions.
In addition, rich environmental elements, including trees, guardrails, buildings, and roadside infrastructures, are integrated to characterize complex propagation conditions in wide-lane scenarios. 
Owing to its diversity and fidelity, the dataset provides a versatile foundation for U2V sensing–communication studies, supporting a range of modeling and learning tasks.


\item A novel multi-modal data-driven 3D-SPADE algorithm is proposed for 3D scatterer prediction for the first time, which explores the mapping relationship between physical environment and electromagnetic space.
To address the information loss caused by single-view point clouds, 3D-SPADE innovatively fuses point clouds from transmitter (Tx) and receiver (Rx), followed by noise removal and spatial voxelization to unify multi-modal data formats. 
A novel customized deep network based on the 3D U-Net architecture is designed to accurately predict the 3D spatial distribution of scatterers by leveraging multi-modal features, thereby enabling the exploration of the mapping relationship between LiDAR point clouds and scatterer distributions.

\item A multi-modal channel model based on 3D scatterer prediction is constructed for the first time to support 6G U2V intelligent communications. 
The mapping relationship between physical environment and electromagnetic space is investigated to mitigate excessive dependence on data while enhancing physical interpretability.
By virtue of the voxel-based representation inherent in the proposed 3D-SPADE algorithm, scatterers are clustered with the 3D voxel grid as the basic spatial unit, within which dynamic and static clusters are distinguished to represent corresponding environmental objects. 
As LiDAR point clouds evolve, these clusters are jointly tracked via 3D-SPADE, enabling accurate characterization of channel non-stationarity and consistency with reduced computational complexity.

\item 
Through comprehensive quantitative and qualitative evaluations, including statistical analysis of cluster prediction and cluster visualization, the effectiveness of the proposed 3D-SPADE is validated.
The corresponding key channel statistical characteristics, such as the time autocorrelation function (TACF) and Doppler power spectral density (DPSD), are derived and analyzed.
These characteristics from the simulation experiments are consistent with ray-tracing simulations and exhibit better agreement than with the standardized model and the state-of-the-art model, demonstrating the necessity of exploring the mapping relationship and the effectiveness of the proposed model.
\end{enumerate}

The remainder of this paper is organized as follows. Section II introduces the mixed sensing-communication integration simulation dataset. The 3D-SPADE algorithm, which can explore the mapping relationship between physical environment and electromagnetic space, is developed in Section III. Section IV describes the proposed multi-modal intelligent U2V channel model. Key channel statistical properties are given in Section V. In Section VI, the simulation result is presented and is further compared with the ray-tracing results, the standardized model, and the state-of-the-art model. Finally, Section VII concludes.

\section{INTELLIGENT SENSING-COMMUNICATION INTEGRATED MEASUREMENT CAMPAIGN OF U2V COMMUNICATION}
\begin{figure*}[!t] 
    \centering
    \includegraphics[width=0.8\textwidth, keepaspectratio]{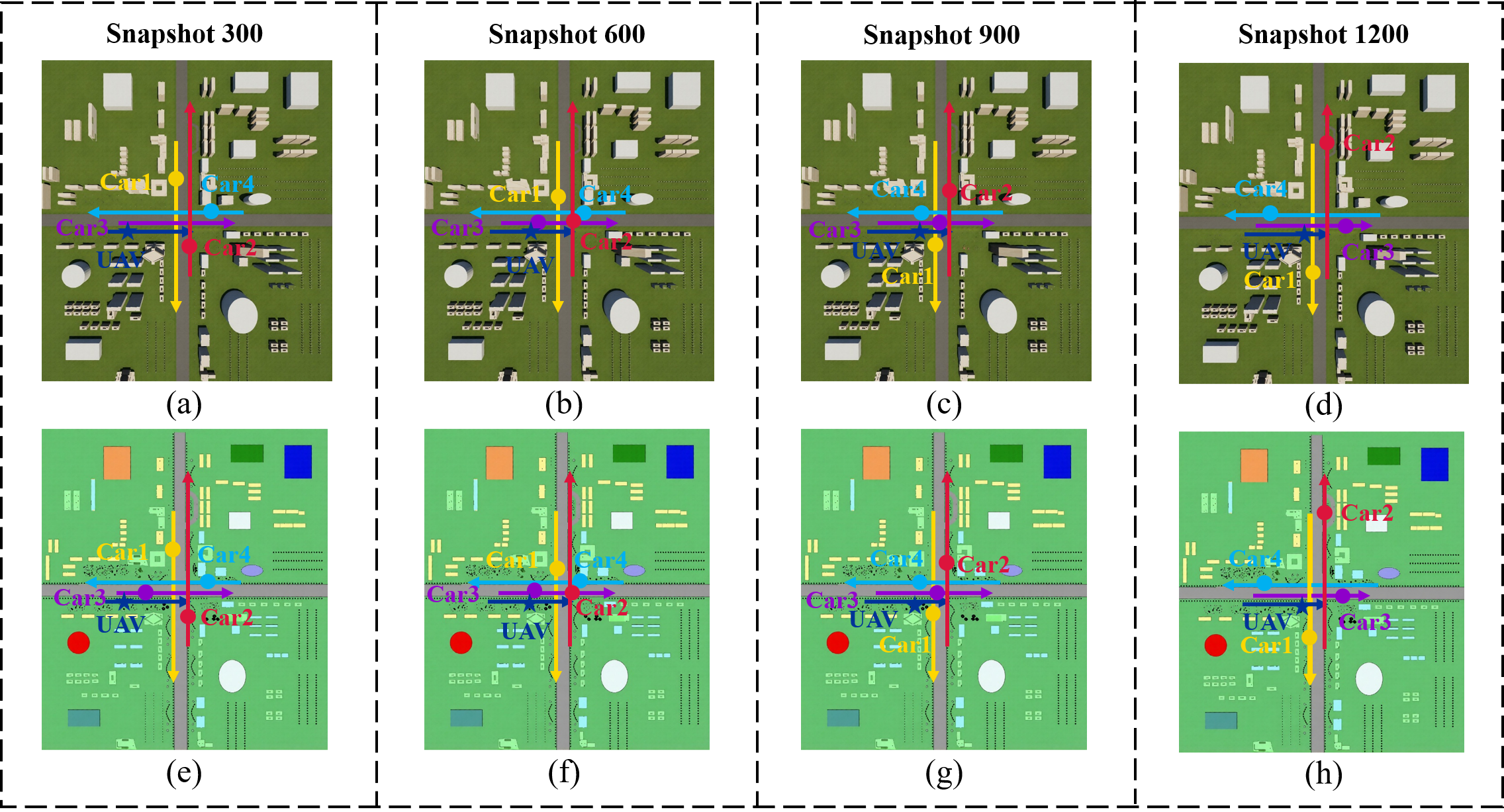}
     \caption{
    Dynamic scenarios at Snapshots 300, 600, 900, and 1200 under low VTD. The top row (a–d) and bottom row (e–h) correspond to the environments in AirSim and Wireless InSite, respectively. In each snapshot, the star represents the real-time position of the UAV, solid circles denote the positions of ground vehicles, and the arrows indicate the moving direction of each object.
    } 
    \label{fig:scenarios} 
\end{figure*}
To explore the mapping relationship between physical environment and electromagnetic space, a wide-lane scenario in electromagnetic space, and those in the physical environment are constructed and precisely aligned, thereby a high-fidelity mixed sensing-communication integration simulation U2V dataset is constructed.

\subsection{Scenario Construction and Precise Alignment}
Wireless InSite and AirSim are utilized as the primary simulation platforms in this study to generate and acquire communication and sensing data.
Wireless InSite leverages ray-tracing (RT) technology to model radio wave propagation and wireless communication systems \cite{wi}.
AirSim is an open-source platform constructed on Unreal Engine to acquire high-fidelity sensing data \cite{airsim}. 
In accordance with the characteristics of the wide-lane U2V scenario, the environment is configured by incorporating representative roadside and urban elements, including trees, guardrails, and buildings, together with other factors that influence U2V propagation, to support comprehensive sensing and communication data acquisition.
To guarantee the precise alignment of the scenario, the physical environment is first modeled in AirSim with high-fidelity geometric and semantic descriptions, and is then imported into Wireless InSite to construct the corresponding electromagnetic space.
During this process, the geometry, object dimensions, material properties, and time-varying motion trajectories of all entities are strictly preserved, establishing a one-to-one correspondence between the physical and electromagnetic domains.
As a result, each LiDAR point cloud snapshot is exactly synchronized with the corresponding RT-based propagation snapshot, following the paradigm in \cite{m3sc}.

\subsection{Parameter Settings and Data Collection}
To ensure sufficient data volume for comprehensive analysis and robust validation of the proposed algorithms, the number of snapshots is set to 1500 per VTD condition, with a time interval of 0.01 s between snapshots.
To enhance the diversity and comprehensiveness of the dataset, the wide-lane scenario includes different VTDs, UAV heights, and communication links. 
The numbers of vehicles in high, medium, and low VTDs are 30, 20, and 10, respectively. 
The UAV height is set to 60~m and 20~m.
Meanwhile, each vehicle and UAV is equipped with a LiDAR device and a communication unit. 
The LiDAR device in AirSim is configured with 16 channels and operates at a scanning frequency of 10 Hz.
The communication unit in Wireless InSite functions within the standard mmWave frequency band, with a carrier frequency of 28 GHz and a communication bandwidth of 2 GHz. 
Each communication unit is equipped with both a Tx and a Rx, where the number of antennas at both the Tx and Rx is set to $M_T = M_R = 1$.
For each VTD condition, UAV is selected as the Tx, and four vehicles moving in different directions are selected as the Rx, forming four communication links.
After constructing the scenario and setting the parameters, the electromagnetic propagation scatterers for a single U2V communication link are obtained by leveraging RT technology, and LiDAR point cloud data is acquired via AirSim, collectively constructing a comprehensive dataset for subsequent analysis and processing.

The 3D coordinates of the UAV and ground vehicles are updated synchronously at each snapshot, enabling continuous 3D motion of the UAV and multiple vehicles. 
Fig.~\ref{fig:scenarios} visualizes the dynamic evolution of the low VTD wide-lane scenario at Snapshots 300, 600, 900, and 1200, where the one-to-one precise alignment between the physical environment in AirSim and the electromagnetic simulation space in Wireless InSite is guaranteed throughout the entire dynamic process.

Through the above scenario configuration and parameter settings, a large-scale and diverse U2V dataset is constructed, featuring multi-modal sensing and communication data, continuous 3D mobility, and multiple communication links under various traffic and flight conditions.
Specifically, the dataset covers four representative scenarios, including high-, medium-, and low-VTD conditions at a UAV height of 60 m, as well as a high-VTD condition at a UAV height of 20 m, where U2V data are collected along four typical movement directions in each scenario. For each scenario, 1500 snapshots of LiDAR point clouds and corresponding communication data are acquired.


\section{The 3D-SPADE Algorithm for LiDAR-Aided Scatterer Representation}
To enable accurate channel modeling in highly dynamic and complex U2V scenarios, it is essential to characterize the relationship between the physical environment and the electromagnetic space. However, this relationship is inherently complex and cannot be described by explicit analytical expressions.

To address this issue, this section proposes the 3D-SPADE algorithm. 
3D-SPADE decodes the physical environment encoded in LiDAR point clouds, and characterizes the electromagnetic propagation space via the spatial distribution of scatterers.
Instead of directly mapping sensing data to channel parameters, 3D-SPADE learns the mapping between these two representations, thereby providing a physically meaningful intermediate representation for channel modeling.

The proposed 3D-SPADE algorithm consists of two main components. First, a preprocessing module constructs structured representations of the physical environment and the electromagnetic space using LiDAR point clouds and 3D scatterers, respectively. 
Second, a deep neural network is designed to learn the voxel-wise mapping between these two representations and estimate the corresponding scatterer distribution.
This two-stage progressive design, with preprocessing as the prerequisite and mapping learning as the core, is detailed in Section III-A and Section III-B, respectively.
\subsection{Construction of Physical and Electromagnetic Representations}
\begin{figure*}[!t] 
    \centering
    \includegraphics[width=0.8\textwidth, keepaspectratio]{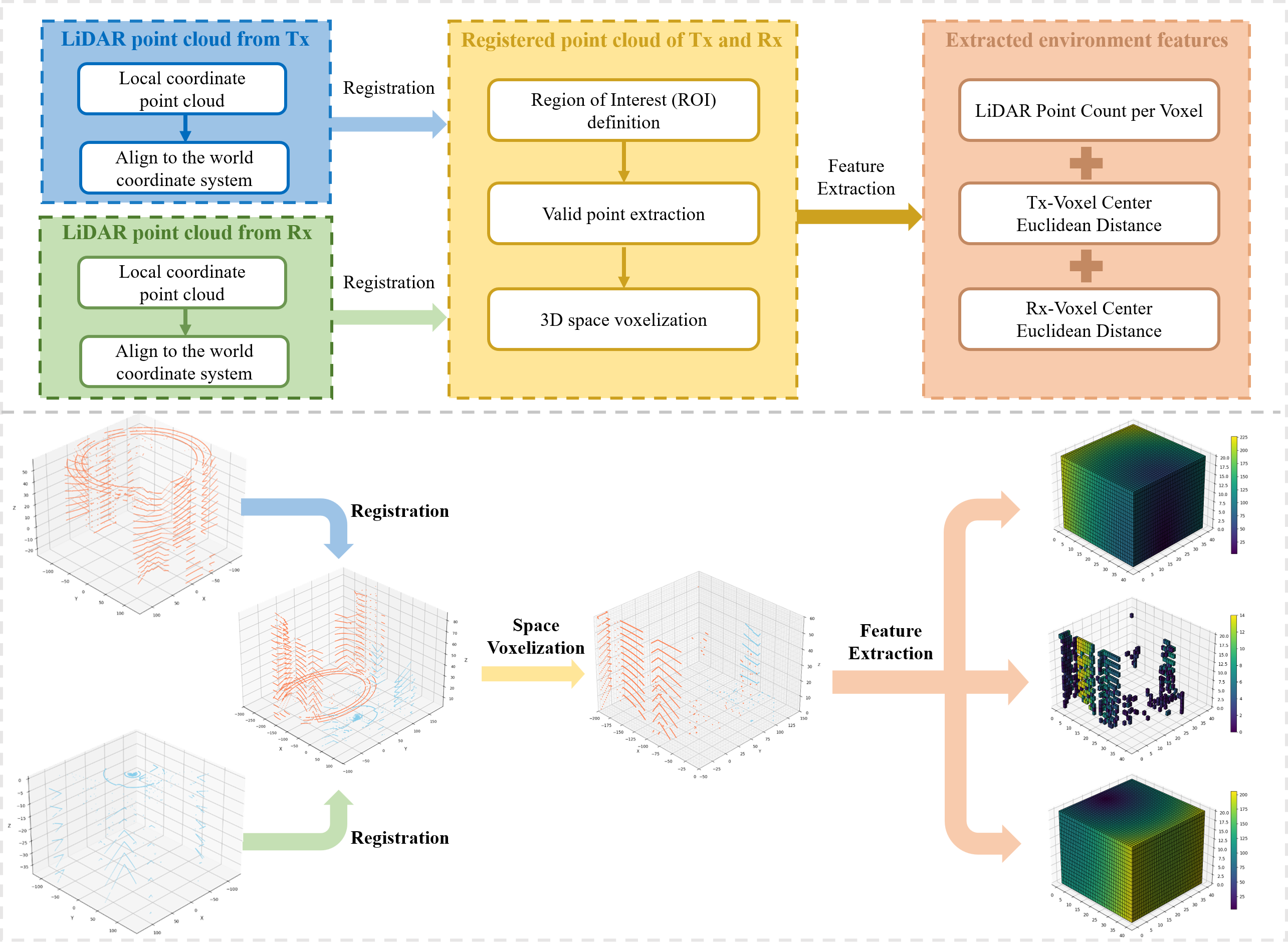}
    \caption{Processing pipeline from LiDAR point clouds to network input features and output label. } 
    \label{fig:processing} 
\end{figure*}

In U2V communication scenarios, LiDAR point clouds provide high-resolution 3D descriptions of the surrounding physical environment, whose geometric characteristics critically influence electromagnetic space. 
However, raw LiDAR point clouds are inherently irregular, unstructured, and viewpoint-dependent, making them unsuitable for direct use in learning-based modeling and channel analysis. Therefore, a dedicated preprocessing procedure is required to extract physical environment information related to electromagnetic space while converting raw sensing data into a structured and spatially consistent representation.
As illustrated in Fig.~\ref{fig:processing}, the proposed preprocessing pipeline consists of coordinate transformation and registration, valid point cloud extraction, spatial voxelization, and environment feature extraction. 

\subsubsection{Coordinate Transformation and Registration}
LiDAR point clouds are recorded in the local coordinate system of each sensor. 
To enable joint processing of LiDAR data collected at the Tx and Rx, all point clouds are transformed into a unified world coordinate system. 
This transformation also facilitates the registration of multi-view point clouds, thereby mitigating spatial information loss caused by single-view observations and enabling the joint representation of aerial and ground-level structures.

Let the position of a sensing platform at time $t$ be denoted by $[x_s(t),y_s(t),z_s(t)]$, and let $[x_l(t),y_l(t),z_l(t)]$ denote a LiDAR point in the local sensor coordinate system. 
Given the platform heading angle $\theta$ with respect to the $x$-axis, the corresponding world-coordinate representation of the LiDAR point is expressed as
\begin{equation}
\mathbf{L}(t)=
\begin{bmatrix}
x_s(t)-x_l(t)\cos\theta+y_l(t)\sin\theta \\
y_s(t)-y_l(t)\cos\theta+x_l(t)\sin\theta \\
z_s(t)-z_l(t)
\end{bmatrix}
\end{equation}
where $\theta$ is the corresponding heading angle, equal to $\theta_{\mathrm{T}}$ for the Tx and $\theta_{\mathrm{R}}$ for the Rx.

Using the above transformation, the LiDAR point sets collected at the Tx and Rx at time $t$, denoted by $P_{\mathrm{Tx}}(t)$ and $P_{\mathrm{Rx}}(t)$, are expressed in the same global reference frame and can be directly merged as
\begin{equation}
P_{\mathrm{registered}}(t)=P_{\mathrm{Tx}}(t)\cup P_{\mathrm{Rx}}(t).
\end{equation}

This simple yet effective registration strategy yields a unified geometric representation of the surrounding environment, serving as the basis for subsequent voxelization and feature extraction.

\subsubsection{Spatial Voxelization}
After coordinate transformation and registration, spatial voxelization is performed to convert the irregular LiDAR point clouds into a structured 3D representation, which enables reliable voxel-wise feature extraction, subsequent mapping relationship learning, and low-complexity scatterer clustering. 
Specifically, the physical environment is partitioned into a regular voxel grid, where each voxel corresponds to a small cubic region in space. 
This voxelization process transforms unstructured points into a form suitable for voxel-wise learning and spatial feature extraction.

Before voxelization, a region of interest (ROI) is defined by the spatial bounds $[X_{\min}, X_{\max}]$, $[Y_{\min}, Y_{\max}]$, and $[Z_{\min}, Z_{\max}]$. 
LiDAR points outside this region are discarded. 
In addition, to suppress propagation-irrelevant clutter caused by ground reflections, a height threshold $H_t$ is applied, and points below this threshold are removed. 
The resulting valid point set is denoted as $P_{\mathrm{valid}}$.

The ROI is then uniformly partitioned into a voxel grid of size $g_x \times g_y \times g_z$, with voxel resolutions given by
\begin{equation}
L_{x/y/z} = \frac{{X/Y/Z}_{\max} - {X/Y/Z}_{\min}}{g_{x/y/z}}.
\end{equation}

Each valid LiDAR point is mapped to a unique voxel according to its spatial coordinates,
\begin{equation}
i_{x/y/z} = \left\lfloor \frac{{x/y/z}_p - {X/Y/Z}_{\min}}{L_{x/y/z}} \right\rfloor.
\end{equation}
where \(x_p, y_p, z_p\) are the 3D spatial coordinates of a valid LiDAR point \(p\).

Through this process, the registered point cloud is transformed into a regular voxelized representation.
Moreover, the voxel grid provides a natural spatial abstraction for scatterer clustering, where scatterers within the same voxel can be treated as the same cluster without relying on explicit clustering algorithms. 
More importantly, this voxel-based representation unifies the spatial formats of LiDAR point clouds and scatterer distributions, thereby facilitating consistent voxel-wise learning between sensing data and electromagnetic representations.

\subsubsection{Environment Feature Extraction}
After voxelization, voxel-level environmental features are extracted to characterize the physical factors that influence electromagnetic space and scatterer distribution. 
Rather than using high-dimensional or redundant descriptors, a compact set of geometry- and propagation-related features is designed to capture the essential spatial characteristics of the environment.

Specifically, three voxel-level features are considered. The first feature is the number of LiDAR points within each voxel, which reflects geometric density and implicitly indicates the presence and complexity of physical objects. 
Regions with higher point density are more likely to contribute to electromagnetic scattering.

The second feature is the Euclidean distance between the voxel center and the Tx, which captures the distance-dependent interaction between scatterers and the transmitted signal. Voxels closer to the Tx tend to have a stronger influence on signal reflection and scattering. The distance from voxel $V_i$ to the Tx is computed as
\begin{equation}
\mathbf{D}_{\text{Tx}}(V_i)=\sqrt{(x_i^c-x_{\mathrm{Tx}})^2+(y_i^c-y_{\mathrm{Tx}})^2+(z_i^c-z_{\mathrm{Tx}})^2},
\end{equation}
where $[x_i^c,y_i^c,z_i^c]$ denotes the center of voxel $V_i$ and $[x_{\text{Tx}},y_{\text{Tx}},z_{\text{Tx}}]$ denotes the location of Tx.

The third feature is the Euclidean distance from the voxel center to the Rx, defined similarly as
\begin{equation}
\mathbf{D}_{\text{Rx}}(V_i)=\sqrt{(x_i^c-x_{\mathrm{Rx}})^2+(y_i^c-y_{\mathrm{Rx}})^2+(z_i^c-z_{\mathrm{Rx}})^2}.
\end{equation}
where $[x_{\text{Rx}},y_{\text{Rx}},z_{\text{Rx}}]$ denotes the location of Rx.

This feature characterizes the contribution of each voxel to the received signal and complements the Rx distance in describing propagation geometry.

By combining voxel density and Tx/Rx distance information, the resulting multi-channel voxel representation jointly encodes environmental geometry and propagation relevance. 
This compact yet informative feature design facilitates effective learning of the mapping between LiDAR-based environment perception and electromagnetic scatterer distribution.

\subsection{Representation Mapping via the Deep Neural Network}
\begin{figure*}[!t] 
    \centering
    \includegraphics[width=0.8\textwidth, height=9cm, keepaspectratio]{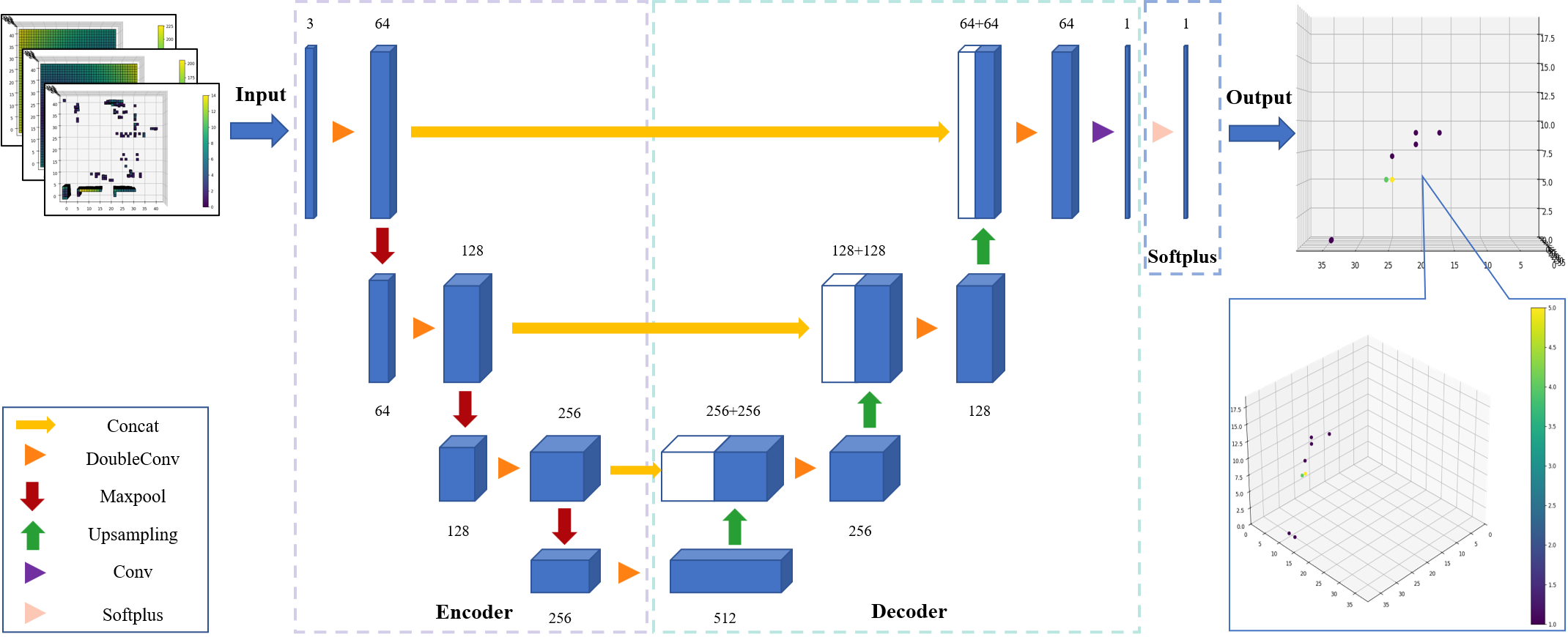}
    \caption{Customized DNN network based on the 3D U-Net architecture.} 
    \label{fig:3D U-Net} 
\end{figure*}

The mapping relationship between the physical environment representation derived from LiDAR sensing and the electromagnetic scatterer representation is highly nonlinear and difficult to characterize using analytical models. 
To address this challenge, a customized deep neural network (DNN) based on 3D U-Net architecture is developed, which is specifically designed to learn the voxel-wise mapping between the two domain representations in a data-driven manner, as illustrated in Fig.~\ref{fig:3D U-Net}.

Based on the voxelized representations constructed in the preprocessing stage, the network takes structured LiDAR-based environmental features as input and outputs the corresponding spatial distribution of scatterers. 
Rather than directly predicting channel parameters, the network focuses on learning an intermediate scatterer representation that preserves physical interpretability and facilitates subsequent channel modeling.

\subsubsection{Input and Output Representations}
Based on the representations constructed in the preprocessing stage, the input to the network is a three-channel 3D voxel grid of size $g_x \times g_y \times g_z$. 
Each voxel encodes the corresponding environmental features derived from LiDAR point clouds, forming a structured input representation for voxel-wise learning.
The network output is a single-channel voxel grid with the same spatial resolution as the input, where each voxel value represents the estimated number of electromagnetic scatterers within the corresponding spatial cell. 
Ground-truth scatterer distributions are obtained from high-fidelity channel propagation data and voxelized using the same grid, enabling direct voxel-wise supervision for representation mapping.

\subsubsection{Network Architecture and Physical Constraints}
To realize voxel-wise mapping between physical environment and electromagnetic space representations while preserving spatial consistency, an encoder–decoder architecture based on 3D convolutional operations is adopted. 
3D U-Net acts as the core backbone of the deep neural network in the 3D-SPADE algorithm, which is inherently suitable for processing the 3D voxelized volumetric data in the U2V scatterer spatial distribution prediction task and excels at capturing multi-scale spatial features of scatterers in wide-lane scenarios \cite{3dunet}. 
Its customized architecture aligns perfectly with the voxel-based environmental features generated in the preprocessing stage, ensuring accurate voxel-wise scatterer prediction, and the voxelized prediction results can directly support scatterer clustering by treating each voxel as an independent cluster, thus significantly reducing the computational complexity of subsequent channel modeling while enhancing its physical interpretability.

Each encoder block consists of two successive 3D convolutional layers followed by batch normalization and nonlinear activation, which together form a double convolution module. 
This operation is defined as
\begin{equation}
\begin{aligned}
\operatorname{DoubleConv}(X) &= \operatorname{ReLU}\left( \operatorname{BN}\left( \operatorname{Conv3d}\left( \right. \right. \right. \\
&\quad \operatorname{ReLU}\left( \operatorname{BN}\left( \operatorname{Conv3d}(X) \right) \right) \left. \left. \left. \right) \right) \right).
\end{aligned}
\end{equation}

For the $l$-th encoder block, given the input feature map ${X}^{(l)}$, the operations are as follows
\begin{equation}
Z^{(l)} = \text{DoubleConv}\left(X^{(l+1)}\right)
\end{equation}
\begin{equation}
X^{(l+1)} = \text{MaxPool3d}\left( Z^{(l)} \right)
\end{equation}
where ${X}^{(l+1)}$ is the feature map after the maxpooling operation, and the output of the encoder block is passed to the next block for further processing.

The decoder mirrors the encoder structure and restores spatial resolution through a sequence of upsampling operations. 
At each decoder stage, a 3D transposed convolution is first applied to upsample the feature map,
\begin{equation}
U^{(l)} = \operatorname{ConvTranspose3d}\left( {X}^{(l+1)} \right),
\end{equation}
which is then concatenated with the corresponding encoder feature map via skip connections,
\begin{equation}
C^{(l)} = [U^{(l)}, Z^{(l)}]
\end{equation}
This fusion enables the network to retain fine-grained spatial details that are critical for accurate voxel-level scatterer estimation. The concatenated features are subsequently refined using the double convolution operation,
\begin{equation}
{X}^{(l)} = \operatorname{DoubleConv}\left( C^{(l)} \right).
\end{equation}

After the final decoder stage, a $1 \times 1 \times 1$ 3D convolution is applied to project the decoded features into the output space. To enforce physical consistency, a Softplus activation function is used at the output layer,
\begin{equation}
O = \operatorname{Softplus}\left( \operatorname{Conv3d}\left({X}^{(1)} \right) \right),
\end{equation}
which guarantees non-negative predictions of scatterer counts. This constraint reflects the inherent physical property that the number of electromagnetic scatterers within a voxel cannot be negative. Mathematically, the Softplus function is defined as:
\begin{equation}
\operatorname{Softplus}(x) = \ln\left(1 + e^x\right)
\end{equation}

By integrating voxel-wise learning with explicit physical constraints, the proposed network effectively learns a physically meaningful mapping from LiDAR-based environmental representations to electromagnetic scatterer distributions.
\section{A MULTI-MODAL INTELLIGENT U2V CHANNEL MODELING VIA 3D-SPADE}
\begin{figure}[!t] 
    \centering 
    \includegraphics[width=0.5\textwidth]{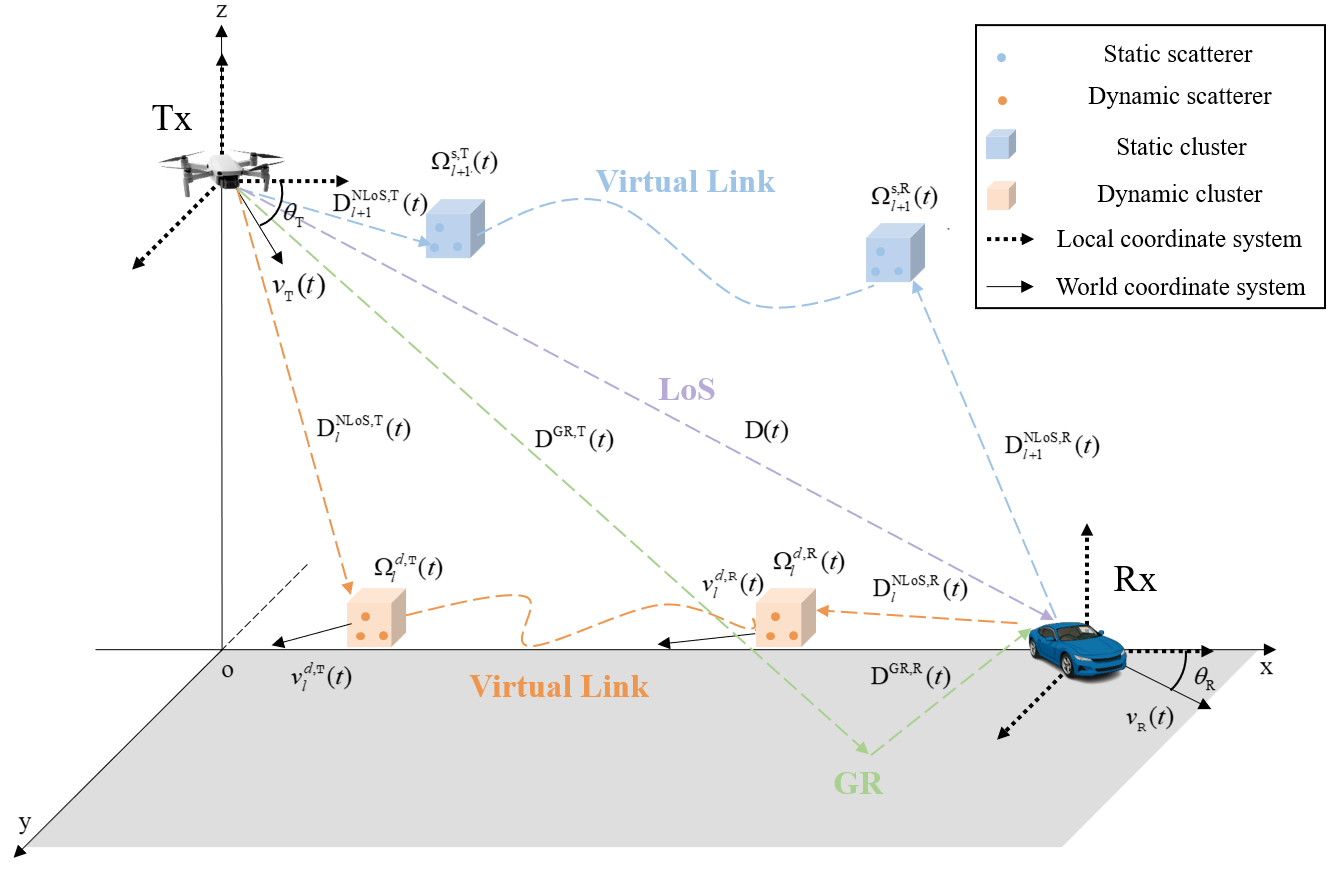} 
    \caption{Geometry of the proposed multi-modal intelligent U2V channel model.} 
    \label{fig:u2v model}
\end{figure}

The proposed multi-modal U2V channel model for the wide-lane scenario is illustrated in Fig.~\ref{fig:u2v model}.
The Tx is the UAV, and the Rx is a moving vehicle equipped with a communication device and a LiDAR device.
The channel impulse response (CIR) \(h(t,\tau)\) of the transmission from Tx to
Rx includes the line-of-sight (LoS) component, the non-LoS (NLoS) component via dynamic and static clusters, and the ground reflection component.
Using LiDAR point clouds and the mapping relationship between the physical environment and electromagnetic space through 3D-SPADE, the spatial distribution of scatterers is predicted. 
Based on the predicted scatterers, clustering is performed, and dynamic and static clusters are classified. 
Additionally, the non-stationarity and consistency of the U2V channel are modeled.

\subsection{Processing of Predicted Scatterer}
Before channel modeling, the predicted scatterers need to be processed.
Given that the model yields non-negative continuous values, but the physical representation of scatterers requires non-negative integers, a rounding strategy based on a threshold is employed. 

Specifically, a threshold $\tau$ is set to perform rounding, and the final number of scatterers $\hat{Y}$ is calculated as follows
\begin{equation}
\hat{Y}= \begin{cases} 
\lfloor O \rfloor + 1 & \text{if } (O- \lfloor O\rfloor) \geq \tau, \\
\lfloor O\rfloor  & \text{otherwise}.
\end{cases}
\label{eq:output}
\end{equation}
where $\lfloor \cdot \rfloor$ denotes the floor operator, i.e., rounding down to the nearest integer.

The next step is to define and standardize the scatterer clusters, which are formed by grouping the voxel-wise scatterers output via the proposed 3D-SPADE algorithm.
Each 3D voxel grid, which serves as the basic input and output unit of the 3D-SPADE network, represents the minimum discrete unit of the 3D propagation space. All scatterers predicted within the same voxel are uniformly defined as a single independent scatterer cluster.
The geometric center of the corresponding voxel grid is set as the spatial center of each cluster, providing a unified coordinate reference for subsequent U2V channel parameter calculation.

After scatterer clustering, the classification of dynamic and static clusters is essential for accurately modeling highly dynamic and complex environments. 
The dynamic clusters, which represent moving objects, typically have lower heights, while static scatterers, such as buildings or trees, have higher heights \cite{dynamic_static_height}.
To distinguish between dynamic and static clusters, a threshold $H_c$ based on height is applied. 
Specifically, if the height of a cluster is below the threshold $H_c$, it is classified as a dynamic cluster. 
Conversely, if the height exceeds the threshold $H_c$, the cluster is classified as a static scatterer. 
Once the cluster is classified as either dynamic or static based on the threshold $H_c$, the scatterers within that cluster are also classified accordingly.
This classification allows the model to effectively represent the time-varying nature of dynamic scatterers and the consistent presence of static scatterers.

\subsection{Framework of the Proposed Channel Model}
After the processing of the predicted scatterers, the scatterer number, distance, angle, power, and status for each cluster can be determined.
$\Omega(t)$ denotes the Ricean factor, $\eta^{\text{GR}}(t)$ and $\eta^{\text{NLoS}}(t)$ are the power ratios of the ground reflection component and NLoS component via dynamic/static clusters.
Meanwhile, it is satisfied that
\(
\eta^{\text{GR}}(t) + \eta^{\text{NLoS}}(t) = 1,
\) as described in \cite{channelmodeling}.

\subsubsection{LoS Component}
The distance vector between Tx and Rx at the initial time is $\mathbf{D}(t_0)$. The distance between Tx and Rx at time $t$ is computed by
\begin{equation}
    \mathbf{D}(t) = \mathbf{D}(t_0) + \int_{t_0}^{t} \mathbf{v}_{\text{R}}(t) \, dt - \int_{t_0}^{t} \mathbf{v}_{\text{T}}(t) \, dt
\end{equation}
where $\mathbf{v}_{\text{R}}(t)$ and $\mathbf{v}_{\text{T}}(t)$ are the velocities of Rx and Tx.

To control the validity of the channel within a specific time range, a window function $Q(t)$ is employed, which can be denoted by
\begin{equation}
 Q(t) = 
\begin{cases} 
1, & t_0 \leq t \leq T_0, \\
0, & \text{otherwise}.
\end{cases}
\end{equation}

The complex channel gain of the LoS component can be represented as
\begin{equation}
h^{\text{LoS}}(t) = Q(t) \exp\left[ j2\pi \int_{t_0}^{t} f^{\text{LoS}}(t) dt + j \varphi^{\text{LoS}}(t) \right] 
\end{equation}
where the Doppler frequency \( f^{\text{LoS}}(t) \), phase shift \( \varphi^{\text{LoS}}(t) \), as well as delay \( \tau^{\text{LoS}}(t) \) are computed by
\begin{equation}
f^{\text{LoS}}(t) = \frac{\langle \mathbf{D}(t), \mathbf{v}_{\text{R}}(t) - \mathbf{v}_{\text{T}}(t) \rangle}{\lambda \| \mathbf{D}(t) \|}
\end{equation}
\begin{equation}
\varphi^{\text{LoS}}(t) = \varphi_0 + \frac{2\pi}{\lambda} \| \mathbf{D}(t) \|
\end{equation}
\begin{equation}
\tau^{\text{LoS}}(t) = \frac{\| \mathbf{D}(t) \|}{c}
\end{equation}
where \( \langle \cdot , \cdot \rangle \) denotes the inner product, \( \varphi_0 \) is the initial phase
shift, \( \lambda \) is the carrier wavelength.

\subsubsection{ NLoS Component via Clusters} 
The complex channel gain via the $l$-th cluster \(\mathcal{C}_l\) can be denoted by
\begin{equation}
\begin{aligned}
h_{l}^{\text{NLoS}}(t) &= Q(t) N_l(t)\sqrt{P_{l}^{\text{NLoS}}(t)} \exp\left\{ j2\pi \left[ \int_{t_0}^{t} f_{l}^{\text{NLoS,T}}(t) dt \right. \right. \\
& \left. \left. + \int_{t_0}^{t} f_{l}^{\text{NLoS,R}}(t) dt \right] + j \varphi_{l}^{\text{NLoS}}(t) \right\}
\end{aligned}
\end{equation}
where $N_l(t)$ represents the number of scatterers in the \( l \)-th cluster \(\mathcal{C}_l\), and \( P_{l}^{\text{NLoS}}(t) \) represents the normalized cluster power and can be denoted by \cite{powerpath}
\begin{equation}
P_l^{\text{NLoS}}(t) = \exp\left( -\xi\tau_l^\text{NLoS}(t) - \eta \right) 10^{-\frac{Z}{10}}
\end{equation}
where $\xi$ and $\eta$ denote the delay-related parameters of the path through the clusters, $\tau_l^\text{NLoS}(t)$ denotes the delay of the path through the $l$-th cluster \(\mathcal{C}_l\).
The parameter $Z$ follows Gaussian distribution \(\mathcal{N}\left( 0, \sigma_{E}^2 \right)\).

The Doppler frequency is computed by
\begin{equation}
f_{l}^{\text{NLoS,T/R}}(t) = \frac{\langle \mathbf{D}_l^{\text{NLoS,T/R}}(t), \mathbf{v}_{\text{T/R}}(t) - \mathbf{v}_l(t) \rangle}{\lambda \| \mathbf{D}_l^{\text{NLoS,T/R}}(t) \|}
\end{equation}
where \( \mathbf{D}_{l}^{\text{NLoS, T/R}}(t) \) represents the distance between the Tx/Rx and the \( l\)-th cluster \(\mathcal{C}_l\).  $\mathbf{v}_l(t)$ denotes the velocity of the $l$-th cluster at time $t$. For dynamic clusters, a non-zero value is assigned to $\mathbf{v}_l(t)$ based on the movement of the cluster. Conversely, for static clusters, $\mathbf{v}_l(t)$ is set to zero, indicating no motion of the cluster.

The delay can be written by
\begin{equation}
\tau_{l}^{\text{NLoS}}(t) = \frac{\| \mathbf{D}_{l}^{\text{NLoS,T}}(t) \| + \| \mathbf{D}_{l}^{\text{NLoS,R}}(t) \|}{c} + \tilde{\tau}_l(t)
\end{equation}
where \( \tilde{\tau}_l(t) \) denotes the abstracted delay of the virtual link in the $l$-th cluster \(\mathcal{C}_l\), which follows the exponential distribution. 

The phase shift is denoted by
\begin{equation}
\begin{split}
\varphi_{l}^{\text{NLoS}}(t) = & \, \varphi_0 + \frac{2\pi}{\lambda} \left[ \| \mathbf{D}_{l}^{\text{NLoS,T}}(t) \| +
\| \mathbf{D}_{l}^{\text{NLoS,R}}(t) \| \right. \\
& \left. + c \tilde{\tau}_l(t) \right].
\end{split}
\end{equation}

\subsubsection{Ground Reflection Component}
The complex channel gain of the ground reflection component can be obtained by
\begin{equation}
\begin{aligned}
h^{\text{GR}}(t) & = Q(t) \sqrt{P^{\text{GR}}(t)} \exp\left\{ j2\pi \left[ \int_{t_0}^{t} f^{\text{GR,T}}(t) dt \right. \right. \\
& \left. \left. + \int_{t_0}^{t} f^{\text{GR,R}}(t) dt \right] + j \varphi^{\text{GR}}(t) \right\}
\end{aligned}
\end{equation}
where \( P^{\text{GR}}(t) \), \( f^{\text{GR, T/R}}(t) \), \(\varphi^{GR}(t)\), and \( \tau^{\text{GR}}(t) \) are power, Doppler frequency at transceiver, phase, and delay, respectively.

With the above derivations, the CIR of the proposed multi-modal intelligent U2V channel model can be formulated as
\begin{multline}
h(t,\tau) = \frac{1}{\sqrt{\Omega(t) + 1}} \Bigg\{ 
\sqrt{\Omega(t)} \, h^{\mathrm{LoS}}(t) \delta\left[\tau-\tau^{\mathrm{LoS}}(t)\right] \\
+ \sqrt{\eta^{\mathrm{NLoS}}(t)} \sum_{l=1}^{N(t)} h_{l}^{\mathrm{NLoS}}(t) \delta\left[\tau-\tau_{l}^{\mathrm{NLoS}}(t)\right] \\
+ \sqrt{\eta^{\mathrm{GR}}(t)} h^{\mathrm{GR}}(t) \delta\left[\tau-\tau^{\mathrm{GR}}(t)\right] \Bigg\}
\end{multline}
where $N(t)$ represents the number of clusters at time $t$. 
\subsection{Capturing of Channel Non-Stationarity and Consistency}
For highly dynamic 3D U2V mmWave communication scenarios, the wireless channel exhibits distinct non-stationary characteristics in space, time, and frequency domains, while maintaining consistency in spatial and temporal dimensions. 
To accurately characterize these properties, the proposed model fully leverages the 3D environment perception capability of LiDAR and the scatterer distribution prediction ability of the 3D-SPADE algorithm to realize precise modeling of space-time-frequency non-stationarity, while guaranteeing the space-time consistency of the channel model.
\subsubsection{Space Domain Non-Stationarity}
Different from traditional 2D ground communication scenarios, U2V communications involve 3D arbitrary motion trajectories of the UAV and ground vehicles, leading to prominent space domain non-stationarity. To capture this property, the proposed model leverages the 3D voxelization and voxel-wise scatterer prediction capability of the 3D-SPADE algorithm. The 3D physical environment is discretized into regular 3D voxel grids, and the 3D-SPADE algorithm realizes accurate prediction of scatterer distribution in each voxel. This voxel-wise modeling framework can explicitly characterize the differences of scatterer density, type, and scattering characteristics in different 3D spatial regions, thus fully modeling the space-domain non-stationarity of U2V channels.
\subsubsection{Time Domain Non-Stationarity}
For highly dynamic U2V scenarios, the relative positions between the Tx and Rx, as well as the surrounding scatterer distribution, change continuously over time, resulting in significant time domain non-stationarity. 
The proposed model captures this property via the time-varying LiDAR input of the 3D-SPADE algorithm.
The LiDAR point clouds collected by the Tx and Rx at different time instants are distinct, which reflect the real-time dynamic changes of the physical propagation environment. 
Based on the time-varying LiDAR input, the 3D-SPADE algorithm outputs dynamically updated 3D scatterer distributions, thereby fully capturing the time domain non-stationarity of the U2V channel.
\subsubsection{Frequency Domain Non-Stationarity}
The deployment of mmWave technology in U2V communication channels amplifies the importance of accounting for frequency non-stationarity.
Therefore, in the proposed model, a frequency-dependent factor $\chi$ is incorporated into the time-varying transfer function to mimic the frequency-dependent variations in path gain.
By making the Fourier transform with $h(t,\tau)$ with respect to $\tau$, the time-varying transfer function is derived, as shown in the equation
\begin{equation}
H^{'}(t,f) = \int_{-\infty}^{\infty} h(t,\tau) \exp(-j2\pi f \tau) \mathrm{d}\tau.
\end{equation}

The time-varying transfer function is denoted as
\begin{multline}
H(t,f) = \frac{1}{\sqrt{\Omega(t) + 1}} \Bigg\{ 
\sqrt{\Omega(t)} \, h^{\mathrm{LoS}}(t) \exp\left[-j2\pi f \tau^{\mathrm{LoS}}(t)\right] \\
+ \sqrt{\eta^{\mathrm{NLoS}}(t)} \left(\frac{f}{f_c}\right)^\chi 
\sum_{l=1}^{N(t)} h_{l}^{\mathrm{NLoS}}(t) \exp\left[-j2\pi f \tau_{l}^{\mathrm{NLoS}}(t)\right] \\
+ \sqrt{\eta^{\mathrm{GR}}(t)} \left(\frac{f}{f_c}\right)^\chi 
h^{\mathrm{GR}}(t) \exp\left[-j2\pi f \tau^{\mathrm{GR}}(t)\right] \Bigg\}
\end{multline}
where $\chi$ is the frequency-dependent parameter that relies on the environment \cite{frequency_environment}.
\subsubsection{Spatial Consistency}
In the physical propagation environment, the scatterer distribution and channel characteristics maintain inherent spatial consistency between adjacent 3D spatial regions. The proposed model captures this property via the 3D spatial feature extraction mechanism of the 3D-SPADE algorithm.
Specifically, the customized 3D U-Net backbone is designed to extract the spatial correlation of scatterer distributions between adjacent voxel grids. 
By leveraging its multi-scale receptive fields, the model ensures that the predicted scatterer density and scattering characteristics evolve smoothly and logically across the 3D spatial domain, ensuring the necessary spatial consistency.
\subsubsection{Temporal Consistency}
For the U2V propagation environment, the physical scenario only has minimal changes in adjacent time instants with small time intervals, which means the channel characteristics should maintain temporal continuity without abrupt jumps. 
The proposed model guarantees this temporal-domain consistency via two core designs. 
First, the LiDAR point clouds collected at adjacent time instants only have slight variations, thereby ensuring that the scatterer distribution predicted by the proposed 3D-SPADE algorithm maintains a smooth transition in the time dimension.
Second, the scatterer clustering based on 3D voxel grids ensures that the spatial positions of most predicted clusters remain unchanged at adjacent time instants, with only the number of scatterers in partial clusters changing slightly.
This design fully guarantees the temporal-domain consistency of the proposed model.
\section{CHANNEL STATISTICAL PROPERTIES}
The corresponding key channel statistical properties of the multi-modal U2V channel model for 6G intelligent sensing-communication integration are derived.
\subsection{Time-Frequency Correlation Function}
Based on the time-varying transfer function, the TF-CF is defined as
\begin{equation}
\xi(t, f; \Delta t, \Delta f) = \mathbb{E}[H^*(t, f) H(t + \Delta t, f + \Delta f)]
\end{equation}
where $\mathbb{E}[\cdot]$ denotes the expectation operator and $(\cdot)^*$ represent the complex conjugate operation, respectively. 
Moreover, the TF-CF of LoS components, NLoS components resulting from static and dynamic clusters, and ground reflection components can be calculated similarly to \cite{tfcf}. Based on the TF-CF, the time auto-correlation function (TACF) and the frequency correlation function (FCF) can be derived by setting $\Delta f=0$ and $\Delta t=0$, respectively.

\subsection{Doppler Power Spectral Density}
The DPSD can be derived by taking the Fourier transform of the TACF \cite{dpsd}, namely
\begin{equation}
\Theta(t; f_D) = \int_{-\infty}^{+\infty} \xi(t; \Delta t) e^{-j2\pi f_D \Delta t} \text{d}(\Delta t)
\end{equation}
where $\xi(t ; \Delta t)$ is the TACF and $f_{D}$ is the Doppler frequency.


\section{SIMULATIONS AND ANALYSIS RESULTS} 
This section validates the proposed 3D-SPADE algorithm and the resulting multi-modal U2V channel model through comprehensive simulations and analyses. 
First, the network training procedure and scatterer prediction performance are evaluated using quantitative metrics and qualitative visualizations. 
Then, key channel statistical properties derived from the proposed model are analyzed to demonstrate its capability in capturing channel non-stationarity. 
Finally, the simulation results are compared with RT-based results, as well as with standardized and state-of-the-art channel models, to verify the accuracy and generality of the proposed approach.

\subsection{Network Training and Validation}
\subsubsection{Overview of the Dataset}The simulation U2V dataset under each VTD contains 6000 samples, generated from four pairs of transceivers over 1500 snapshots.
Each sample consists of three-channel LiDAR point cloud data encoded into a regular 3D voxel grid as the network input, where each channel encodes a distinct geometric feature of the physical environment, and the spatial distribution of effective electromagnetic scatterers, encoded into a matched single-channel 3D voxel grid as the output.
For each VTD, the dataset is partitioned into three subsets: training, validation, and test sets, with a distribution ratio of 3:1:1.
Following training on the training set, the performance of the model is evaluated, and hyperparameters are optimized on the validation set.
Subsequently, the performance testing is conducted on the network using the test set.

\subsubsection{Network Configuration and Hyperparameters}Table~\ref{tab:hyper-parameters}
shows hyperparameters for the customized network design,
including the size of the LiDAR point cloud feature voxel grid,
the size of the scatterer voxel grid, and the architecture details of the
3D-SPADE network first. 
Additionally, the hyperparameters used during the fine-tuning process are also listed in Table~\ref {tab:hyper-parameters}.

\subsubsection{Simulation Results and Analysis of Scatterer Prediction}
To verify the performance of the developed 3D-SPADE algorithm, three standard classification metrics are adopted: precision, recall, and F1.  
According to Eq.~\eqref{eq:output}, the predicted number of scatterers in the $i$-th voxel grid is $\hat{y_i}$, and the ground truth is $y_i$, then the following quantities can be defined 
\begin{align}
\mathrm{TP}&=\bigl|\{\,i: y_i \neq 0,\ \hat{y}_i \neq 0\,\}\bigr|\\
\mathrm{FP}&=\bigl|\{\,i: y_i=0,\ \hat{y}_i \neq 0\,\}\bigr|\\
\mathrm{FN}&=\bigl|\{\,i: y_i \neq 0,\ \hat{y}_i=0\,\}\bigr|\\
\mathrm{TN}&=\bigl|\{\,i: y_i=0,\ \hat{y}_i=0\,\}\bigr|
\end{align}
\begin{equation}
\mathrm{Precision}=\frac{\mathrm{TP}}{\mathrm{TP}+\mathrm{FP}}
\end{equation}
\begin{equation}
\mathrm{Recall}=\frac{\mathrm{TP}}{\mathrm{TP}+\mathrm{FN}}
\end{equation}
\begin{equation}
\mathrm{F1}=\frac{2\,\mathrm{Precision}\cdot \mathrm{Recall}}{\mathrm{Precision}+\mathrm{Recall}}.
\end{equation}

In scatterer prediction, precision reflects the reliability of predicted scatterer voxels by measuring the false-alarm level, while recall characterizes the completeness of detection by quantifying the miss-detection behavior. 
The F1-score jointly considers precision and recall and captures the overall balance between false positives and false negatives. 

\begin{figure*}[!t]
    \centering 
    \includegraphics[width=1.0\textwidth]{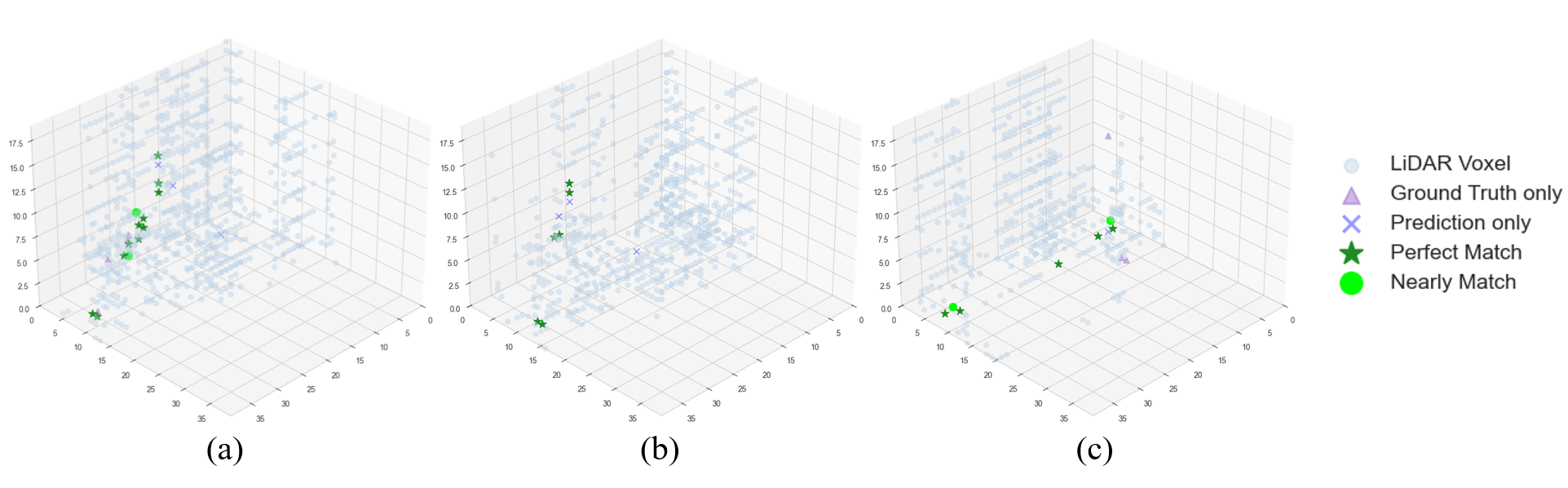}  
    \label{subfig:a} 
    \vspace{-0.4cm}  
    \caption{Comparison between the predicted clusters and ground truth under high VTD across different snapshots. (a) Comparison at snapshot 150. (b) Comparison at snapshot 300. (c) Comparison at snapshot 450.}
    \label{fig:comparison} 
\end{figure*}

Table~\ref{tab:accuracy} presents the precision, recall, and F1 of scatterer prediction under low, medium, and high VTDs at the wide-lane scenarios, as well as for different UAV heights.
As vehicle density increases and height changes, the evaluation metrics remain stable and maintain a high level, indicating that the developed 3D-SPADE retains robustness while performing well under a complex U2V scenario.
This enables the model to effectively explore the mapping relationship between physical environment and electromagnetic space in highly dynamic and complex wide-lane U2V scenarios.
Fig.~\ref{fig:comparison} presents the comparison between the predicted clusters and ground truth under high VTD across different snapshots under the wide-lane U2V scenario.
The light blue points in the background of each subplot represent 3D voxel grids containing valid LiDAR point clouds, which characterize the geometric structure of the physical propagation environment.
Purple triangles indicate scatterers present only in the ground truth, whereas blue crosses correspond to scatterers predicted only by the network. 
Green stars denote exactly matched voxels, and green circles represent nearly matched voxels whose predicted positions are within a close spatial vicinity of the ground truth.
It can be clearly observed that effective electromagnetic scatterers are concentrated in regions with dense LiDAR point clouds, which confirms the strong correlation between LiDAR-derived environmental features and electromagnetic scatterer distribution. Meanwhile, most scatterers are either exactly matched or predicted with high spatial proximity to the ground truth, with only a small number of mismatched points, indicating that the proposed 3D-SPADE algorithm is capable of accurately predicting the spatial distribution of scatterers under the complex U2V scenario.

\begin{table}[!t]
\centering
\renewcommand{\arraystretch}{1.0}
\setlength{\tabcolsep}{6pt}
\caption{Hyper-parameters for Network Architecture and Training}
\label{tab:hyper-parameters}
\begin{tabular}{
>{\centering\arraybackslash}m{1.0cm}
>{\centering\arraybackslash}m{4.7cm}
>{\centering\arraybackslash}m{1.9cm}
}
\toprule
\textbf{Category} & \textbf{Parameter} & \textbf{Value} \\
\midrule

\multirow{8}{*}{\makecell{\textbf{Network}\\\textbf{Design}}}
& Input voxel grid size $[C_{\mathrm{in}},D,H,W]$   
& $[3,40,40,20]$ \\

& Output voxel grid size $[C_{\mathrm{out}},D,H,W]$ 
& $[1,40,40,20]$ \\

& Encoder kernel size                                
& $(3,3,3)$      \\

& Decoder kernel size                                
& $(3,3,3)$      \\

& Pooling / Unpooling kernel size                    
& $(2,2,2)$      \\

& Skip connections                                   
& Yes            \\

& Output convolution kernel                          
& $(1,1,1)$      \\

& Output activation                                  
& Softplus       \\
\midrule

\multirow{5}{*}{\textbf{Training}}
& Batch size     
& 4                 \\

& Learning rate  
& $1.0\times10^{-3}$ \\

& Epochs         
& 120               \\

& Optimizer      
& Adam              \\

& Loss function  
& MSE Loss          \\
\bottomrule
\end{tabular}
\end{table}

\begin{table}[!t]
\centering
\renewcommand{\arraystretch}{1.0}
\setlength{\tabcolsep}{6pt}
\caption{Accuracy of Scatterer Voxel Grid Under Low, Medium, and High VTDs}
\label{tab:accuracy}
\begin{tabular}{
>{\centering\arraybackslash}m{3.4cm}
>{\centering\arraybackslash}m{1.3cm}
>{\centering\arraybackslash}m{1.3cm}
>{\centering\arraybackslash}m{1.3cm}
}
\toprule
\textbf{Scenario conditions} & \textbf{Precision} & \textbf{Recall} & \textbf{F1} \\
\midrule
Low VTD, height = 60 m    & 93.77\% & 86.57\% & 90.02\% \\
Medium VTD, height = 60 m & 95.74\% & 83.11\% & 89.75\% \\
High VTD, height = 60 m   & 94.52\% & 84.38\% & 89.16\% \\
High VTD, height = 20 m   & 90.91\% & 93.26\% & 91.14\% \\
\bottomrule
\end{tabular}
\end{table}

\subsection{Model Simulation}
\begin{figure}[!t] 
    \centering 
    \includegraphics[width=0.4\textwidth]{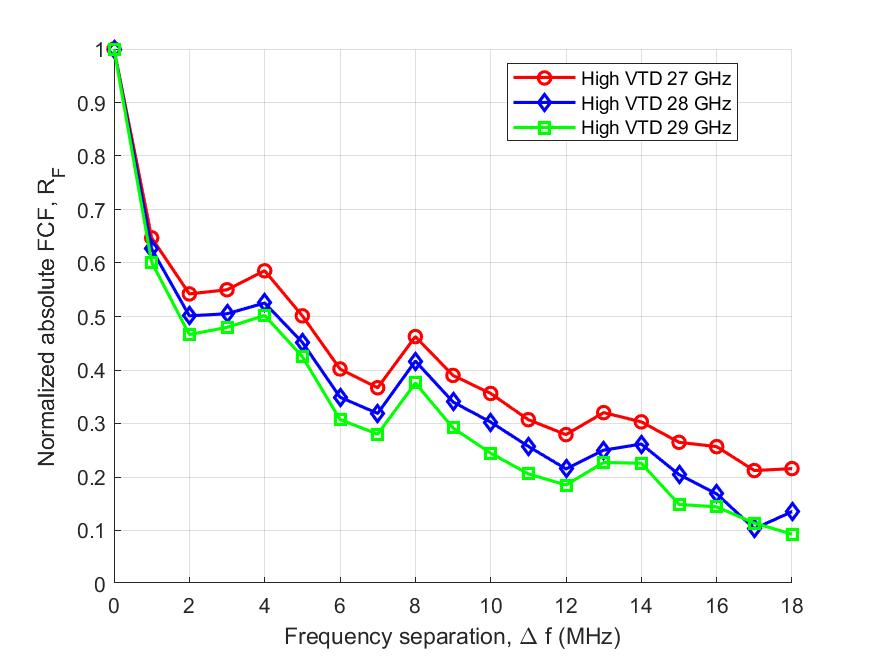} 
    \caption{Absolute normalized FCF for different frequencies under the high VTD.}
    \label{fig:fcf} 
\end{figure}

Fig.~\ref{fig:fcf} shows the absolute normalized FCF for different frequencies under the high VTD at the wide-lane U2V scenario.
It can be observed that the FCF values decrease as the frequency separation increases, demonstrating the capturing of frequency non-stationarity of the proposed channel model.
Meanwhile, the 27 GHz curve remains at the top for the majority of the time, while the 29 GHz curve consistently stays at the bottom. 
This is primarily due to the more complex channel characteristics at higher frequencies, where increased path loss and susceptibility to blockage result in lower frequency correlation. 
These factors cause higher frequencies to experience more rapid signal attenuation and variability in the channel, leading to a reduced correlation with the lower frequencies.

\begin{figure}[!t]
    \centering 
    \includegraphics[width=0.4\textwidth]{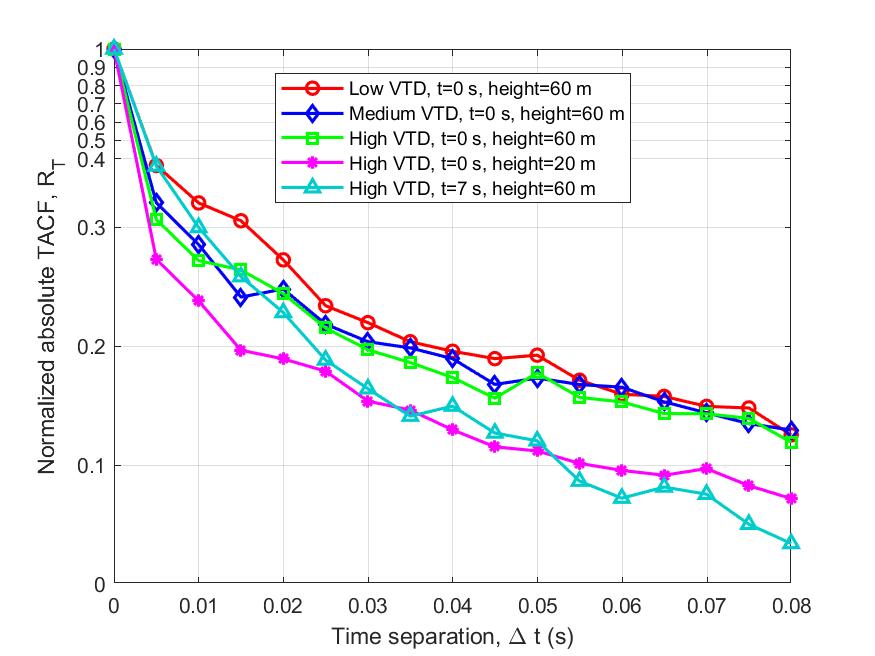} 
    \caption{Absolute normalized TACF under different VTDs, time instants, and UAV heights.} 
    \label{fig:tacf} 
\end{figure}

Fig.~\ref{fig:tacf} presents the absolute normalized TACF values under different VTDs, times, and UAV heights. 
For the same scenario configuration, the TACF exhibits noticeable variations across different time instants \(t\), which evidences the time non-stationary behavior captured by the proposed model.
Furthermore, when the UAV is deployed at a height of 60~m, the TACF curves under different VTDs exhibit marginal differences within the same time separation. 
The reason for this behavior is that ground vehicles are lower than the UAV height and therefore have a limited influence on the main propagation paths.
This behavior is further supported by the comparison between different UAV heights, where a lower UAV height results in reduced TACF values due to the stronger influence of ground objects that increase the complexity of the propagation paths.
Moreover, for the same VTD, the TACF values exhibit differences across different time intervals.
This variation is attributed to the different positions of the Tx and Rx, with environmental factors affecting the scatterers, which in turn directly influence the TACF.

\subsection{Model Verification}
The deterministic channel model based on the RT technology is of high accuracy. Specifically, relying on the Wireless InSite simulation platform, the RT technology is exploited to obtain the CIR in wide-lane U2V communication scenarios. 
Based on the CIR, the generality of the proposed channel model is verified.

\begin{figure}[!t]
    \centering %
    \includegraphics[width=0.4\textwidth]{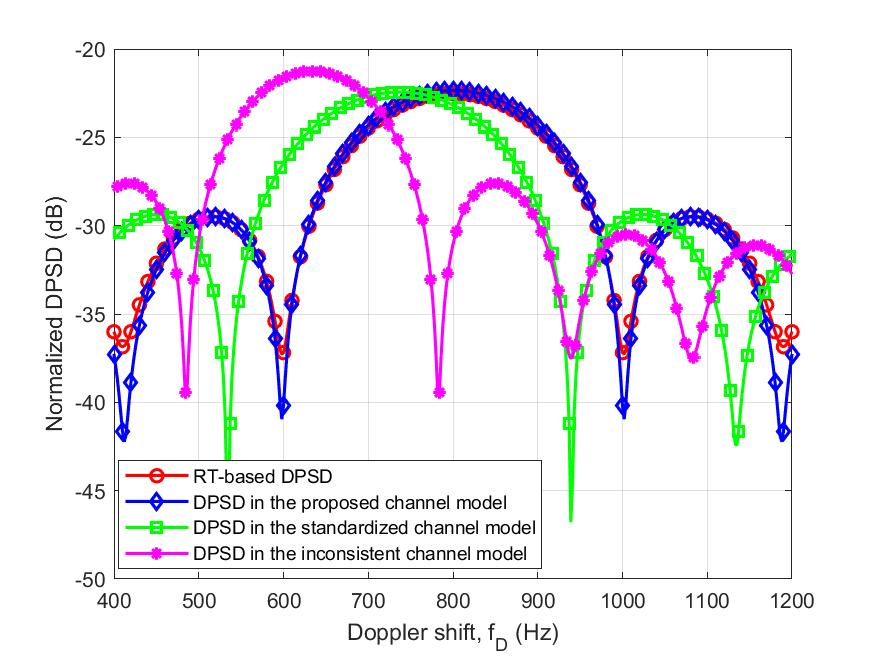} 
    \caption{Comparison of DPSD among the RT-based result, the proposed channel model, the standardized model, and the inconsistent channel model.} 
    \label{fig:comparison_dpsd} 
\end{figure}

The RT-based DPSD is derived by processing the RT-based CIR under the high VTD.
Simulated DPSD for high VTD based on the proposed channel model is also calculated and compared with the RT-based DPSD results. 
To validate the generality of the proposed model, the DPSD calculated by the standardized model \cite{standardized} and the inconsistent model \cite{inconsistent} are also used for comparison. 
As shown in Fig.~\ref{fig:comparison_dpsd}, the simulation results can match well with RT-based results, outperform those of the standard and inconsistent models, validating the effectiveness of the proposed model.
The standardized model tends to rely on generic parameterizations and simplified Doppler representations, which may yield sparse effective Doppler components and thus introduce artificial spectral notches. 
Although the inconsistent model explicitly separates dynamic and static scattering contributions, its Doppler spectrum still depends on predefined statistical assumptions and power allocations; when these assumptions deviate from the actual wide-lane U2V conditions, noticeable mismatches in spectral structure and notch patterns can occur.

\begin{figure}[!t] 
    \centering 
    \includegraphics[width=0.4\textwidth]{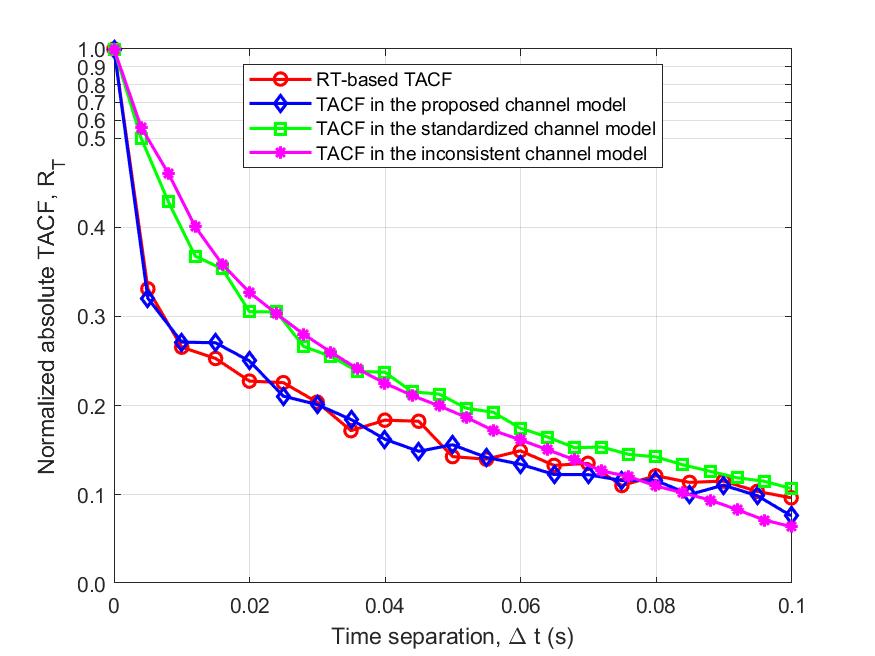} 
    \caption{Comparison of TACF among the RT-based result, the proposed channel model, the standardized model, and the inconsistent channel model.} 
    \label{fig:comparison_tacf}
\end{figure}

The RT-based CIR under high VTD is processed to obtain the RT-based TACF. In Fig.~\ref{fig:comparison_tacf}, the RT-based TACF is compared with the TACF derived from the proposed model, the TACF from the standardized model, and the TACF based on the channel model that does not consider consistency. It is evident that the TACF produced by the proposed model, which accounts for time non-stationarity, closely aligns with the RT-based TACF. 
This excellent agreement demonstrates that the proposed model accurately characterizes the impact of scatterers in highly dynamic and complex U2V communication scenarios. 
In contrast, the TACF generated by the standardized model fails to match the RT-based result, as the model lacks precise information regarding scatterer positions and quantities.
Furthermore, the inconsistent model, which employs parameters that do not align with the actual U2V scenario, results in an overly smoothed TACF that fails to accurately capture the RT-based decay behavior. 
A similar issue is observed in the standardized model, both of them relying on statistical distributions of scatterers rather than precise scatterer positions, leading to significant discrepancies in the TACF.


\section{CONCLUSIONS}
This paper has proposed a multi-modal intelligent U2V channel model for 6G intelligent sensing-communication integration, based on 3D scatterer prediction. 
A large-scale and diverse U2V dataset has been constructed, featuring multi-modal sensing and communication data, continuous 3D mobility, and multiple communication links under various traffic and flight conditions.
Based on the constructed dataset, the 3D-SPADE algorithm has been developed to accurately predict the spatial distribution of scatterers using LiDAR point clouds. 
The mapping relationship between the physical environment and electromagnetic space explored by 3D-SPADE is further exploited for scatterer clustering and dynamic–static classification, thereby reducing modeling complexity while improving accuracy.
By integrating 3D-SPADE, clustering, and cluster classification, a multi-modal intelligent U2V channel model has been constructed.
As LiDAR point clouds vary over time, dynamic and static clusters evolve via 3D-SPADE, enabling precise modeling of channel non-stationarity and consistency. 
Simulation results demonstrate that, in the wide-lane scenario with varying VTDs and UAV heights, the proposed 3D-SPADE consistently achieves high scatterer occupancy detection performance within the voxel grid. In particular, under favorable configurations, recall reaches 93.26\%, and precision reaches 95.74\%, highlighting the reliability of 3D-SPADE.
Key channel statistical properties, such as DPSD and TACF, have been derived and analyzed. 
The close fit between simulation results and ray-tracing results, with better agreement than the standardized model and inconsistent model, has demonstrated the necessity of exploring the mapping relationship and the effectiveness of the proposed model.
\bibliographystyle{IEEEtran}
\bibliography{references}

\end{document}